\documentclass[prc,superscriptaddress,noshowpacs,unsortedaddress,twocolumn,showpacs,preprintnumbers,amsmath,amssymb]{revtex4-2}

\usepackage[dvipdfmx]{graphicx}
\usepackage{amsmath,amssymb,times}
\usepackage{color}
\usepackage{ulem}
\usepackage{bm}
\usepackage{here}

\def\lsim{~\,\makebox(1,1){$\stackrel{<}{\widetilde{}}$}\,~}

\newcommand{\beq}{\begin{equation}}
\newcommand{\eeq}{\end{equation}}
\newcommand{\bea}{\begin{eqnarray}}
\newcommand{\eea}{\end{eqnarray}}

\newcommand{\bfi}[1]{\mbox{\boldmath $#1$}}

\newcommand{\vK}{{\bfi K}}

\newcommand{\vs}{{\bfi s}}

\newcommand{\vrr}{{\bfi r}}
\newcommand{\vR}{{\bfi R}}

\def\a{\alpha}

\begin{document}

\title{  $^{12}$C+$^{12}$C scattering  as the reference system for  reaction cross section}

\author{Shingo~Tagami}
\affiliation{Department of Physics, Kyushu University, Fukuoka 819-0395, Japan}

\author{Tomotsugu~Wakasa}
\affiliation{Department of Physics, Kyushu University, Fukuoka 819-0395, Japan}

\author{Masanobu Yahiro}
\email[]{orion093g@gmail.com}
\affiliation{Department of Physics, Kyushu University, Fukuoka 819-0395, Japan}

\begin{abstract}
\begin{description}
\item[Background]
In our previous paper, we tested the chiral (Kyushu) folding model for  $^{12}$C+$^{12}$C scattering, 
since the profile function in the Glauber mode is constructed for the system. 
We found that the folding model is reliable for reaction cross sections $\sigma_{\rm R}$
in $30  \lsim E_{\rm lab} \lsim 100 $~MeV and $250  \lsim E_{\rm lab} \lsim 400 $~MeV. 
Accurate data are available for $^{12}$C scattering on  $^{9}$Be, $^{12}$C, $^{27}$Al targets 
in  $30  \lsim E_{\rm lab} \lsim 400 $~MeV. 
\item[Purpose]
We determine matter radius  $r_{m}({\rm exp})$ of $^{12}$C  
from the accurate $\sigma_{\rm R}({\rm exp})$, using the Kyushu $g$-matrix 
folding model.  
\item[Results]
Our result is $r_{\rm m}^{12}({\rm exp}) =2.352 \pm 0.013$~fm for $^{12}$C. 
The model is applied for  the accurate data on $^{12}$C+$^{27}$Al scattering, and yields 
$r_{\rm m}({\rm exp}) =2.936 \pm 0.012$~fm for $^{27}$Al. 
\item[Conclusion]
Our conclusion is  that  $r_{\rm m}({\rm exp}) =2.352 \pm 0.013$~fm agrees with 
$r_{\rm m}({\rm exp}) =2.35 \pm	0.02$~fm determined from interaction cross sections by 
Tanihata {\it et. al.}. 
\end{description}
 \end{abstract}

\maketitle

\section{Introduction and Conclusion}
\label{Sec:Introduction}

{\it Background on experiments and models:} 
We consider that $^{12}$C+$^{12}$C scattering  is the reference system 
for  reaction cross section $\sigma_{\rm R}$, 
since the profile function in the Glauber mode is constructed for the system; for example, 
see Ref.~\cite{Abu-Ibrahim:2007ehr}. 
In our previous paper of Ref.~\cite{PRC.101.014620}, 
 we tested the chiral (Kyushu) folding model~\cite{Toyokawa:2017pdd} for  $^{12}$C+$^{12}$C scattering, 
 and found that the folding model is reliable for reaction cross sections $\sigma_{\rm R}$
in $30  \lsim E_{\rm lab} \lsim 100 $~MeV and $250  \lsim E_{\rm lab} \lsim 400 $~MeV 
by using high-accurate data~\cite{Takechi:2009zz} mentioned below.

Takechi {\it et. al.} measured $\sigma_{\rm R}$ for $^{12}$C scattering on  $^{9}$Be, $^{12}$C, $^{27}$Al targets in  $30  \lsim E_{\rm lab} \lsim 400 $~MeV 
with 2\% errors~\cite{Takechi:2009zz}.
Thus, high-accurate data  are available for  $^{12}$C+ $^{12}$C,  
$^{12}$C+$^{27}$Al scattering  in  $30  \lsim E_{\rm lab} \lsim 400 $~MeV.

Our model is composed of the chiral (Kyushu) $g$-matrix  folding model for scattering~\cite{Toyokawa:2017pdd} 
and D1S-GHFB+AMP~\cite{Tagami:2019svt} for target densities, where D1S-GHFB+AMP stands for Gogny-D1S HFB with the angular momentum projection (AMP). 
The Kyushu $g$-matrix folding model was already applied for p+  $^{208}$Pb scattering~\cite{Tagami:2020bee}. 
The model reproduces the data 
on reactions cross sections $\sigma_{\rm R}$, but not the central values of the data. 
We then determined a skin value  from the data 
by changing $r_{\rm skin}$ of D1S-GHFB+AMP slightly. Our result 
 $r_{\rm skin}$ = $0.278 \pm 0.035$ fm agrees with the recent PREX2 result, i.e.,   
$r_{\rm skin}$ = $0.283\pm 0.071$ fm~\cite{Adhikari:2021phr}.

In Ref.~\cite{Kamimura:1981oxj}, Kamimura constructed the matter density  of  $^{12}$C 
for the ground state by using the 3$\alpha$ RGM model; see Table 2 of his paper.  
The ground state reproduces electron scattering. 
However, the matter radius $r_{\rm m}=2.40$~fm of the 
ground state is slightly larger than the experimental vale $r_{\rm m}({\rm exp}) =2.35 \pm	0.02$~fm~\cite{Tanihata:1988ub} determined from interaction cross sections by 
Tanihata {\it et. al.}.

{\it Aim:}
We use the Kyushu folding model, whenever we calculate $\sigma_{\rm R}$. 
As for $^{12}$C+$^{12}$C scattering, the  $\sigma_{\rm R}$ calculated 
with the phenomenological matter density are compared with 
those with D1S-GHFB+AMP and 3$\alpha$ RGM. 
We determine matter radius  $r_{m}({\rm exp})$ of $^{12}$C  
from the accurate data~\cite{Takechi:2009zz}.

{\it Results:}
 The $\sigma_{\rm R}$ calculated with the phenomenological matter density yield  
 better agreement with the accurate data ~\cite{Takechi:2009zz}
 than those with D1S-GHFB+AMP and the 3$\alpha$ RGM model. 
We then scale the phenomenological proton and neutron  densities and determine  $ r_{\rm m}({\rm exp})$ 
from  the accurate data  by scaling $r_{\rm m}$ under 
the condition that $r_p({\rm scaling})=r_p({\rm exp})$, where $r_{p}({\rm scaling})$ 
($r_{p}({\rm exp})$) denotes 
the scaled  (experimental) proton radius. 
The result thus obtained is  $r_{\rm m}({\rm exp}) =2.352 \pm	0.013~{\rm fm}$. 
The same scaling procedure is used for accurate data~\cite{Takechi:2009zz}  
on  $^{27}$Al + $^{12}$C scattering; 
the result  for $^{27}$Al is $r_m=2.936 \pm  0.012$~fm.  
 The reason why we determine $r_m=2.936 \pm  0.012$~fm for $^{27}$Al  is not shown in 
 Ref.~\cite{Ozawa:2001hb}. 
 
 Our results are summarized in Table  \ref{TW values}.

\begin{table}[htb]
\begin{center}
\caption
{Values of  $r_{\rm p}$, $r_{\rm m}$,  neutron radii $r_{\rm n}$, $r_{\rm skin}$.  
The $r_{\rm p}({\rm exp})$ are determined from the charge radii~\cite{Angeli:2013epw}. 
`Data' shows citations on $\sigma_{\rm R}$. 
The radii are shown in units of fm.  
 }
\begin{tabular}{cccccc}
\hline\hline
  $r_{\rm p}({\rm exp})$ & $r_{\rm m}({\rm exp})$ &  $r_{\rm n}({\rm exp})$ & $r_{\rm skin}({\rm exp})$ & Data \\
\hline
  $^{12}$C~  $2.3272$ & $2.352 \pm 0.013$ & $2.377 \pm	0.027$ & $0.050 \pm 0.027$ & \cite{Takechi:2009zz} \\
  $^{27}$Al~  $2.9483$ & $2.936 \pm 0.012$ & $2.924 \pm 0.023$ & $-0.024 \pm 0.023$ & \cite{Takechi:2009zz} \\
\hline
\end{tabular}
 \label{TW values}
 \end{center} 
 \end{table}

{\it Conclusion:}
Our conclusion is  that  $r_{\rm m}({\rm exp}) =2.352 \pm 0.013$~fm agrees with 
$r_{\rm m}({\rm exp}) =2.35 \pm	0.02$~fm~\cite{Tanihata:1988ub} determined 
from interaction cross sections. 
We also determine  $r_m=2.936 \pm  0.012$~fm for  $^{27}$Al from 
$\sigma_{\rm R}$~\cite{Takechi:2009zz}.

\section{Mehod}
\label{Sec:Method}

 As for the symmetric nuclear matter, Kohno calculated the $g$ matrix 
by using the Brueckner-Hartree-Fock (BHF) method with chiral N$^{3}$LO 2NFs and NNLO 3NFs~\cite{Kohno:2012vj}. 
The framework is applied for positive energies. The resulting non-local chiral  $g$ matrix is localized 
into three-range Gaussian forms by using the localization method proposed 
by the Melbourne group~\cite{von-Geramb-1991,Amos-1994}. 
The resulting local  $g$ matrix is referred to as Kyushu  
$g$-matrix in this paper~\cite{Toyokawa:2017pdd}.

The brief formulation of the double folding model itself is shown below. 
The potential $U$ consists 
of the direct part ($U^{\rm DR}$) and the exchange part ($U^{\rm EX}$):
\bea
\label{eq:UD}
U^{\rm DR}(\vR) \hspace*{-0.15cm} &=& \hspace*{-0.15cm} 
\sum_{\mu,\nu}\int \rho^{\mu}_{\rm P}(\vrr_{\rm P}) 
            \rho^{\nu}_{\rm T}(\vrr_{\rm T})
            g^{\rm DR}_{\mu\nu}(s;\rho_{\mu\nu}) d \vrr_{\rm P} d \vrr_{\rm T}, \\
\label{eq:UEX}
U^{\rm EX}(\vR) \hspace*{-0.15cm} &=& \hspace*{-0.15cm}\sum_{\mu,\nu} 
\int \rho^{\mu}_{\rm P}(\vrr_{\rm P},\vrr_{\rm P}-\vs)
\rho^{\nu}_{\rm T}(\vrr_{\rm T},\vrr_{\rm T}+\vs) \nonumber \\
            &&~~\hspace*{-0.5cm}\times g^{\rm EX}_{\mu\nu}(s;\rho_{\mu\nu}) \exp{[-i\vK(\vR) \cdot \vs/M]}
            d \vrr_{\rm P} d \vrr_{\rm T},~~~~
            \label{U-EX}
\eea
where $\vs=\vrr_{\rm P}-\vrr_{\rm T}+\vR$ 
for the coordinate $\vR$ between P and T. The coordinate 
$\vrr_{\rm P}$ 
($\vrr_{\rm T}$) denotes the location for the interacting nucleon 
measured from the center-of-mass of the projectile (target). 
Each of $\mu$ and $\nu$ corresponds to the $z$-component
of isospin.
The original form of $U^{\rm EX}$ is a non-local function of $\vR$,
but  it has been localized in Eq.~\eqref{U-EX}
with the local semi-classical approximation~\cite{Brieva-Rook-1,Brieva-Rook-2,Brieva-Rook-3} in which
P is assumed to propagate as a plane wave with
the local momentum $\hbar \vK(\vR)$ within a short range of the 
nucleon-nucleon interaction, where $M=A A_{\rm T}/(A +A_{\rm T})$
for the mass number $A$ ($A_{\rm T}$) of P (T).
The validity of this localization is shown in Ref.~\cite{Minomo:2009ds}.

The direct and exchange parts, $g^{\rm DR}_{\mu\nu}$ and 
$g^{\rm EX}_{\mu\nu}$, of the effective nucleon-nucleon interaction 
($g$-matrix) are assumed to depend on the local density
\bea
 \rho_{\mu\nu}=\sigma^{\mu} 
 \left[
 \rho^{\nu}_{\rm T}(\vrr_{\rm T}+\vs/2)+\rho^{\nu}_{\rm P}(\vrr_{\rm P}-\vs/2)
 \right]
\label{local-density approximation}
\eea
at the midpoint of the interacting nucleon pair, where $\sigma^{\mu}$ is the Pauli matrix of a nucleon in P. 

The direct and exchange parts are described by
\begin{align}
&\hspace*{0.5cm} g_{\mu\nu}^{\rm DR}(s;\rho_{\mu\nu}) \nonumber \\ 
&=
\begin{cases}
\displaystyle{\frac{1}{4} \sum_S} \hat{S}^2 g_{\mu\nu}^{S1}
 (s;\rho_{\mu\nu}) \hspace*{0.42cm} ; \hspace*{0.2cm} 
 {\rm for} \hspace*{0.1cm} \mu+\nu = \pm 1 
 \vspace*{0.2cm}\\
\displaystyle{\frac{1}{8} \sum_{S,T}} 
\hat{S}^2 g_{\mu\nu}^{ST}(s;\rho_{\mu\nu}), 
\hspace*{0.2cm} ; \hspace*{0.2cm} 
{\rm for} \hspace*{0.1cm} \mu+\nu = 0 
\end{cases}
\\
&\hspace*{0.5cm}
g_{\mu\nu}^{\rm EX}(s;\rho_{\mu\nu}) \nonumber \\
&=
\begin{cases}
\displaystyle{\frac{1}{4} \sum_S} (-1)^{S+1} 
\hat{S}^2 g_{\mu\nu}^{S1} (s;\rho_{\mu\nu}) 
\hspace*{0.34cm} ; \hspace*{0.2cm} 
{\rm for} \hspace*{0.1cm} \mu+\nu = \pm 1 \vspace*{0.2cm}\\
\displaystyle{\frac{1}{8} \sum_{S,T}} (-1)^{S+T} 
\hat{S}^2 g_{\mu\nu}^{ST}(s;\rho_{\mu\nu}) 
\hspace*{0.2cm} ; \hspace*{0.2cm}
{\rm for} \hspace*{0.1cm} \mu+\nu = 0 ~~~~~
\end{cases}
\end{align}
where $\hat{S} = {\sqrt {2S+1}}$ and $g_{\mu\nu}^{ST}$ are 
the spin-isospin components of the $g$-matrix interaction.
As a way of taking the center-of-mass correction to the phenomenological densities of Ref.~\cite{C12-density}, we use the method of Ref.~\cite{Sumi:2012fr}, since the procedure is quite simple. 
Hamiltonian  based on the 3$\alpha$ RGM model has no 
the center-of-mass coordinate.

In order to deduce the $r_{\rm m}({\rm exp})$, $r_{\rm n}({\rm exp})$, $r_{\rm skin}({\rm exp})$ 
from  measured $\sigma_{\rm R}$~\cite{Takechi:2009zz}, 
we have to scale the proton and neutron densities. 
Now we explain the scaling of density $\rho(\vrr)$.  
We can obtain the scaled density $\rho_{\rm scaling}(\vrr)$ from the original density $\rho(\vrr)$ as
\bea
\rho_{\rm scaling}(\vrr)=\frac{1}{\a^3}\rho(\vrr/\a)
\eea
with a scaling factor
\bea
\a=\sqrt{ \frac{\langle \vrr^2 \rangle_{\rm scaling}}{\langle \vrr^2 \rangle}} .
\eea

\section{Results}
\label{Results} 

\begin{figure}[H]
\begin{center}
 \includegraphics[width=0.5\textwidth,clip]{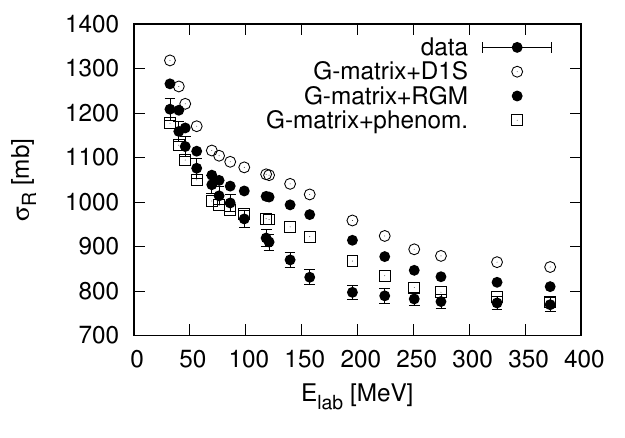}
 \caption{ 
 $E_{\rm lab}$ dependence of reaction cross sections $\sigma_{\rm R}$ 
 for $^{12}$C+$^{12}$C scattering. 
 Open circles stand for the results of the 
 D1S-GHFB+AMP projectile and target  densities. 
 Closed circles correspond to the projectile and target  densities 
 based on the 3$\alpha$ RGM model. 
 Squares denote  the results of the 
 phenomenological projectile and target  densities.
 The data are taken from Refs.~\cite{Takechi:2009zz}. 
   }
 \label{Fig-C+C-kamimura}
\end{center}
\end{figure}

Figure~\ref{Fig-C+C-kamimura} shows $E_{\rm lab}$ dependence 
of reaction cross sections $\sigma_{\rm R}$ for $^{12}$C+$^{12}$C scattering.
The results of the phenomenological projectile and target  densities 
yield better agreement with the data~\cite{Takechi:2009zz} 
than those of the projectile and target  densities 
 based on the 3$\alpha$ RGM model and the D1S-GHFB+AMP projectile and target  densities. Eventually,  the phenomenological projectile and target  densities are best chose for $^{12}$C.

The deviation large in the intermediate energies comes from the $g$-matrix.
We do not introduce any fine-tuning factor $f$, although   in Ref.~\cite{Wakasa:2022ite}
we determined $r_{\rm m}({\rm exp})$ from 
p+$^{12}$C scattering by using $f$.

This indicates that the phenomenological density is a best choice.
Now we determine  $r_{\rm m}({\rm exp})$ from  the data 
in $30 \lsim E_{\rm lab} \lsim 100$~MeV and $250 \lsim E_{\rm lab} \lsim 373$~MeV, scaling the phenomenological projectile and target  
densities. The result is   $r_{\rm m}({\rm exp}) =2.352 \pm 0.013$~fm. 

The same procedure is taken for $^{27}$Al+$^{12}$C scattering 
in $40 \lsim E_{\rm lab} \lsim 87$~MeV and $250 \lsim E_{\rm lab} \lsim 373$~MeV.
Our result is $r_{\rm m}({\rm exp}) =2.936 \pm  0.012$~fm for $^{27}$Al.

\section{Discussions}
\label{Discussion}

When we solve the Shr\"{o}dinger equation having  $U$, the symmetrization between P and T is made for $^{12}$C+$^{12}$C scattering. 
However, as for  $^{27}$Al+$^{12}$C scattering, the symmetrization is not needed.
Table \ref{reference values-a} shows scaling factors, 
$\a_{\rm n}$ and $\a_{\rm p}$, for neutron and proton, respectively. 
In the phenomenological density, the proton density equals the neutron one.  

\begin{table}[htb]
\begin{center}
\caption
{Scaling factors $\a_{\rm n}$ and $\a_{\rm p}$ for neutron and proton.
 }
\begin{tabular}{ccc}
\hline\hline
 & $\a_{\rm n}$ & $\a_{\rm p}$\\
\hline
 $^{12}$C & $1.017$ & $0.996$  \\
 $^{27}$Al  & $0.999 $ & $0.991$ \\
\hline
\end{tabular}
 \label{reference values-a}
 \end{center} 
 \end{table}

In Ref.~\cite{Wakasa:2022ite}, we determined  $r_{\rm m}=2.340 \pm 0.009$~fm 
from $\sigma_{\rm R}$ of p+$^{12}$C scattering by using the fine tuning factor $f$.
In Ref.~\cite{Ozawa:2001hb}, the experimental values on $r_{m}$ are accumulated  from 
$^{4}$He to $^{32}$Mg. Our result  $r_{m}({\rm exp})=2.340 \pm 0.009 $~fm is consistent with  $r_{m}({\rm GSI})=2.35(2)$~fm of Ref.~\cite{Tanihata:1988ub,Ozawa:2001hb}. The present value with no $f$
agrees  with  $r_{m}({\rm GSI})=2.35(2)$~fm. This indicates that the present value is more reliable than $r_{\rm m} =2.340 \pm 0.009$~fm.

Why the $f$ is necessary for p+$^{12}$C but not for 
$^{12}$C+$^{12}$C~? The reason is the following. 
When one determines $r_{\rm m}({\rm exp})$ 
from $\sigma_{\rm R}({\rm exp})$, 
onee use 

\bea
\sigma_{\rm R}({\rm exp})=c_1[r_{\rm m,P}({\rm exp})^2+r_{\rm m,T}({\rm exp})^2]
\label{rm-sigma-cc}
\eea
for $^{12}$C+$^{12}$C scattering, 
\bea
\sigma_{\rm R}({\rm exp})=c_2[r_{\rm m,T}({\rm exp})^2]
\label{rm-sigma-pc}
\eea
for p+$^{12}$C scattering, where the $c_i$ are constants. 
These equations are reliable in the case that P and T are black bodies; 
note that the incident proton is a point-particle. 
The data $\sigma_{\rm R}$ of p+$^{12}$C scattering 
(Fig. 9 of Ref.~\cite{Wakasa:2022ite})
are much smaller than that of$^{12}$C+$^{12}$C scattering 
(Fig.~\ref{Fig-C+C-kamimura}).
 This indicates that  p+$^{12}$C scattering is transparent and 
Eq.~\eqref{rm-sigma-pc} is not correct completely.

\section*{Acknowledgements}
We thank Kamimura and Toyokawa for their contributions.

\bibliography{Folding-v18}

\end{document}